\documentclass[prl,aps,superscriptaddress,twocolumn]{revtex4}
\usepackage{graphicx}
\usepackage{amsmath}
\usepackage{amsfonts}
\usepackage{amssymb}
\usepackage{epsfig}
\usepackage{color}
\usepackage{dsfont}
\usepackage{xcolor}

\newcommand{\numberset}{\mathbb}

\newcommand{\Z}{\numberset{Z}}

\newcommand{\abs}[1]{\left\vert#1\right\vert}

\newcommand{\bra}[1]{\langle#1\vert}
\newcommand{\ket}[1]{\vert#1\rangle}
\newcommand{\Tr}[1]{\text{Tr}\left\{#1\right\}}

\newcommand\braket[2]{\langle#1|#2\rangle}

\begin{document}

\title{Magnetic field effects on electron transport in nanoring with orbital Rashba coupling}

\author{G.~Francica}
\affiliation{CNR-SPIN, I-84084 Fisciano (Salerno), Italy}

\author{P.~Gentile}
\affiliation{CNR-SPIN, I-84084 Fisciano (Salerno), Italy}
\affiliation{Dipartimento di Fisica ``E. R. Caianiello'', Universit\`a di Salerno, I-84084 Fisciano (Salerno), Italy}

\author{M.~Cuoco}
\affiliation{CNR-SPIN, I-84084 Fisciano (Salerno), Italy}
\affiliation{Dipartimento di Fisica ``E. R. Caianiello'', Universit\`a di Salerno, I-84084 Fisciano (Salerno), Italy}

\date{\today}

\begin{abstract}
We study the effects of a Zeeman magnetic field on the electron transport of one-dimensional quantum rings which are marked by electronic states with $d-$orbital symmetry in the presence of spin-orbit and orbital Rashba couplings. By considering phase-coherent propagation, we analyse the geometric Aharonov-Anandan (AA) phase of the channels which is acquired in a closed path, by demonstrating that the orbital polarization can influence the electronic transport when amplitude and magnetic field directions are varied.
We explore all the possible cases for the injection of electrons at various energies in the regime of low electron filling. The magnetic field can allow the selection of only one channel where the transmission is uniquely affected by the AA phase. Conversely, when more orbital channels are involved there is also a dynamical contribution that lead to oscillations in the transmission as the magnetic field is varied. In particular, the transmission is chiral when the energy states are close to the absolute minimum of the energy bands. Instead, when an interference between the channels occurs the orbital and spin contributions tend to balance each other with the increasing of the magnetic field amplitude resulting in a trivial AA phase. This saturation effect does not occur in the high magnetic field regime when orbital and spin properties of the channels exhibit sharp variations with direct consequences on the transport.
\end{abstract}

\maketitle

{\it Introduction. --}
Low dimensional quantum rings allow for phase coherent electron motion within a closed path thus leading to a large variety of fundamental quantum phenonema as for instance the Aharonov-Bohm (AB) \cite{aharanov1959,Webb,Appenzeller,Yacoby,Fuhrer} and Aharonov-Casher (AC) effects \cite{Aharonov1984} and persistent currents \cite{buttiker,levy,landauer,mailly}.
Quantum rings \cite{fomin} are typically realized with semiconducting materials by means of different approaches that include growing quantum dots with ring-like shaped nanostructures, electric field effects, nanolitographic tools, local oxidation techniques, or atom manipulation \cite{atom1,atom2}. Such physical scenario clearly underlines the role of quantum rings as ideal platforms for novel effects and nanoscale electronic engineering.
In particular, when dealing with phase coherent control of the electron spin there are many challenges to face in order to employ the spin degree of freedom and achieve exceptional performances in speed, energy requirements, and functionality.

In this context, novel opportunities are provided by an all-electrical control of the electron spin via electrical means rather than by magnetic fields as it occurs for the AC effect where the electron accumulates a phase when circling in an external electric field or in the presence of inversion asymmetric spin-orbit (SO) interactions.
Remarkably, the electron spin can be also guided when combining spin-orbit coupling in inversion asymmetric semiconducting nanochannels with non-trivial geometric curvature. The manipulation of the electronic states through the corresponding spin geometric phase has been experimentally demonstrated \cite{nagasawa1,nagasawa2}
with the remarkable perspective of achieving topological spin engineering \cite{saarikoski,reynoso}.
The potential of the union of inversion symmetry breaking and nanoscale shaping indeed yields augmenting paths for topological states \cite{saarikoski,reynoso,gentile,ying} and spin-transport \cite{nagasawa2, frustaglia1, koga, bercioux, koga1, Qu}. Such effects have striking geometrical marks as they can strongly depend on the nanoscale shaping especially in narrow spin-orbit coupled semiconducting channels. There, the spin-orbit driving fields act as spatially inhomogeneous geometrical torque controlling both the spin-orientation and its spin-phase through non-trivial spin windings \cite{saarikoski,reynoso,ying}.

Moving beyond conventional semiconductors, transition metal oxides represent alternative material platforms for employing the spin-orbit coupling to steer the electron spin and to combine phase coherent control with other collective phenomena ranging from superconductivity to magnetism in artificial heterostructures with atomically sharp interface. In this context, LaAlO$_3$-SrTiO$_3$ \cite{ohmoto} is a paradigmatic example because a high mobility two-dimensional electron gas (2DEG) forms at the interface of LaAlO$_3$ and SrTiO$_3$, whose carrier density can be tuned by means of top-, side- and back-gating. Remarkably, electric field control of 2DEG transition metal-oxide-based structures have recently enabled the exploration of nanoscale electron quantum transport \cite{caviglia18,bergeal} thus highlighting the prospective of oxide interface in the area of advanced quantum engineering including the potential of achieving topological superconducting phases \cite{schmalian, loder, fukaya1,fukaya2,perroni}. 
Due to the inversion symmetry breaking in the interfacial quantum well a gate-tunable spin-orbit coupling can be generally obtained. A peculiar mark of the oxide 2DEG is that the combination of atomic spin-orbit at the transition element and the orbital dependent (i.e. so called orbital Rashba (OR)) antisymmetric inversion interaction leads to a non-trivial orbital splitting in the reciprocal space. Indeed, while the original Rashba effect for single-band system describes a linear spin splitting and is commonly small in amplitude, in the presence of multi-orbitals a more complex spin-orbital coupled structure arises which can result into a significant splitting.
Indeed, near the $\Gamma$ point, one can have a linear spin-splitting with respect to the momentum for the lowest energy states, but non-linear splittings arise for the intermediate configurations with enhanced anomalies when the filling gets close to band crossings.
The OR interaction generally manifests through mixing of orbitals on neighboring atoms that would not overlap in an inversion symmetric configuration. For instance in oxide 2DEG \cite{gariglio} the OR coupling enables an inversion asymmetric mixing of $d_{xy}$ with $(d_{xz}$,$d_{yz})$ orbitals, and similarly it can also occur in other semiconductors and at the metal's surface.
Taking into account the fundamental characteristics of the spin- and orbital- Rashba couplings it is relevant to ask whether the quantum transport in a ring can manifest imprints that are directly related to the orbital character of the inversion asymmetric coupling of the electronic channels.


To this aim and to assess the role of orbital degrees of freedom for phase coherent quantum control, in the present paper we consider the electron magnetotransport in a single-mode quantum-ring with $d$-orbitals in the presence of atomic spin-orbit and inversion asymmetric OR coupling. The focus is on the role of an applied Zeeman field for both in- and out-of-plane directions.
In general, we find that the electron transmission in the ring for a given injected energy is dependent on the accumulated Aharanov-Anandan (AA) non-adiabatic phase, which in turn directly relates with the solid angles swept by the spin and orbital polarizations.
The AA phase contributes to interference phenomena in the conductance when the ring is symmetrically coupled to two unpolarized leads. The transmission depends on the orbital character of the involved electron channels and a saturation with vanishing amplitude of the AA phase can occur with abrupt transitions when the applied field is strong enough.
An in-plane magnetic field can give a quantization of the AA phase due to the planar symmetry.

{\it The model. --} We consider an effective electronic model that is suitable for transition metal (TM) oxides with perovskite structure where transition metal elements are surrounded by oxygen (O) in an octahedral environment in a tetragonal symmetry.
The model for the 2DEG with broken out-of-plane inversion symmetry has only $t_{2g}$ orbitals, i.e. $\{d_{xy},d_{zx},d_{yz}\}$ at the Fermi level ~\cite{khalsa13,fukaya1,fukaya2}. They are split by the crystal field potential which prefers to have the $xy$ configuration as the lowest one in the orbital hyerarchy.
The electronic connectivity of the $t_{2g}$ bands is highly directional for symmetric TM-O bonds, e.g., an electron in $d_{xy}$-orbital can only hybridize with $p_x$ ($p_y$) states along $y (x)$ directions, respectively, in a square lattice geometry. Other microscopic ingredients include the atomic spin-orbit interaction and the orbital Rashba interaction that couples the momentum to the local orbital angular momentum within the $t_{2g}$ sector.
When considering the electrons moving in a narrow ring of radius $R$ we assume that only the states close to the $\Gamma$ point contribute to the electronic transport and thus the low momentum excitations can be effectively described by an Hamiltonian in the continuum that in the $\{d_{xy},d_{zx},d_{yz}\}$ orbital basis for each spin configuration is expressed as
%
\begin{eqnarray}
\nonumber \mathcal{H}&=& -\frac{1}{R^2}\left(t_2\frac{\mathbf{l}^2}{2}+\left(t_1-t_2\right)l_z^2 \right)\partial_\theta^2 + \Delta_t\left(\frac{{\mathbf{l}}^2}{2}-l_z^2 \right) \\
 && + i \frac{\Delta_{is}}{2R} (l_R \partial_\theta + \partial_\theta l_R)+\lambda_{SO}\mathbf{l}\cdot\boldsymbol{\sigma} -\mathbf B\cdot\boldsymbol \sigma \label{eq.hamiltonian}
\end{eqnarray}

\noindent where $\theta$ is the planar polar angle of the ring, $t_1$ and $t_2$ are the parameters describing the inequivalent effective masses associated to the $(zx,yz)$ and $xy$  orbitals. We point out that for the case of the orbital dependent motion along the ring, the effective mass would a priori depend on the propagation direction and thus on the angle $\theta$. Here, since we focus on the low density regime where the most isotropic $xy$ band is dominating and we are interested in extracting the consequences arising in the phase coherent transport from the interplay of spin and orbital polarizations via the atomic spin-orbit and the orbital Rashba coupling, we neglect the orbital dependence of the effective mass. Hence, $t_1$ and $t_2$ are assumed to be constant.
In the model Hamiltonian (1), $\lambda_{SO}$ is the atomic spin-orbit coupling, $\Delta_{is}$ is the strength of orbital Rashba that breaks inversion, and $\Delta_t$ is the crystal field potential due to flattening of the octahedra. It leads to an energy lowering of the $xy$ orbital with respect to the $(zx,yz)$ states.

The components of $\boldsymbol \sigma$ are the Pauli matrices for the spin operator $\sigma_\alpha$ with $\alpha = x,y,z$, while $l_\alpha$ are the matrices associated with the projected $l=1$ angular momentum
\begin{eqnarray}
\nonumber && l_x = \left(
        \begin{array}{ccc}
          0 & 0 & 0 \\
          0 & 0 & i \\
          0 & -i & 0 \\
        \end{array}
      \right) \quad l_y = \left(
        \begin{array}{ccc}
          0 & 0 & -i \\
          0 & 0 & 0 \\
          i & 0 & 0 \\
        \end{array}
      \right) \\
&& l_z = \left(
        \begin{array}{ccc}
          0 & i & 0 \\
          -i & 0 & 0 \\
          0 & 0 & 0 \\
        \end{array}
      \right)
\end{eqnarray}

\noindent which obey to the commutation relations $[l_\alpha,l_\beta]= -i \epsilon_{\alpha \beta \gamma} l_\gamma$, where $\epsilon_{\alpha \beta \gamma}$ is the antisymmetric tensor, and $l_R$ is the radial component, $l_R=\mathbf l\cdot \hat{R}$.
Concerning the magnetic field $\mathbf B$, we consider only the Zeeman coupling to the spin because the orbital momentum gets locked by the atomic spin-orbit coupling once the symmetry is broken by the external field. We checked that a direct coupling to the orbital angular momentum does not change the outcome of the analysis.

\noindent We start by giving some general considerations on the eigenstates for the case of an applied field $\mathbf B = B_z \hat z$ that is transverse to the orbit of the electrons within the ring. Due to the rotational symmetry, the solution of the Schr\"{o}dinger equation $\mathcal H \ket{\psi_{n \sigma}} = E_{n\sigma} \ket{\psi_{n \sigma}}$ can be expressed as

\begin{equation}\label{eq.sol circle}
\ket{\psi_{n\sigma} (\theta)} = e^{i \left(n+\frac{1}{2}\right)\theta} U_z\left(\theta\right) \ket{\psi_{n\sigma}(0)}
\end{equation}

\noindent where $U_z(\theta)= e^{i (l_z - \frac{\sigma_z}{2})\theta}$ and $n\in \Z$ due to the periodic boundary conditions. Indeed, $J_z = -i\partial_\theta-l_z + \frac{\sigma_z}{2}$ is conserved, and $J_z \ket{\psi_{n\sigma}} = (n+\frac{1}{2})\ket{\psi_{n\sigma}}$.

In particular the state $\ket{\psi_{n\sigma}(0)}$ and the eigen-energy $E_{n\sigma}$ can be directly determined by solving the eigenvalues equation $h_n \ket{\psi_{n\sigma}(0)} = E_{n\sigma} \ket{\psi_{n\sigma}(0)}$ where
\begin{eqnarray}\label{eq.h}
\nonumber h_{n} &=&\frac{1}{R^2}\left(t_2\frac{\mathbf{l}^2}{2}+\left(t_1-t_2\right)l_z^2 \right)\left(l_z-\frac{\sigma_z}{2}+n +\frac{1}{2}\right)^2 \\
\nonumber && +\Delta_t\left(\frac{{\mathbf{l}}^2}{2}-l_z^2 \right)+ \frac{\Delta_{is}}{R}\left[l_x\left(l_z-\frac{\sigma_z}{2}+n +\frac{1}{2}\right) \right. \\&& \left. -i \frac{l_y}{2}\right] + \lambda_{SO} \mathbf{l}\cdot\boldsymbol{\sigma}- B_z \sigma_z
\end{eqnarray}

\noindent We note that for $B_z=0$, as expected, we recover the Kramers degeneracy such that $E_{-n-1 \sigma}=E_{n\sigma}$.

{\it Electron transport and geometric phase. --} The transmission of an electron with energy $E_{in}$ injected in the ring is generally expected to depend on the magnetic field amplitude and direction.
At a given position along the ring identified by the angle $\theta$, a local solution of the time-independent Schr\"{o}dinger equation is expressed as $e^{i k\theta/R } \ket{\chi(k,\theta)}$ where $\ket{\chi(k,\theta)}$ stand for the polarization state at given $\theta$.  In particular the vectors $\ket{\chi(k,\theta)}$ at different points are related through a rotation $\ket{\chi(k,\theta)} = U_z(\theta'-\theta) \ket{\chi(k,\theta')} $. We can thus define $\ket{\chi(k)}\equiv \ket{\chi(k,0)}$, the values of $k$ being determined by solving the linear equation $H_0(k)\ket{\chi(k)}=E_{in}\ket{\chi(k)}$, where

\begin{eqnarray}
\nonumber H_0(k)  &=& \left(t_2\frac{\mathbf{l}^2}{2}+\left(t_1-t_2\right)l_z^2 \right)k^2 + \Delta_t\left(\frac{{\mathbf{l}}^2}{2}-l_z^2 \right) \\
 && - \Delta_{is} l_x k +\lambda_{SO}\mathbf{l}\cdot\boldsymbol{\sigma} - B_z \sigma_z
\end{eqnarray}

\noindent We observe that $I H_0(k)I = H_0(-k)$  where $I$ is an inversion transformation which can be represented through the operator $I=(1-2 l_z^2)\otimes \sigma_z$, from which $E_{\sigma} (k) = E_\sigma (-k)$.

From the continuity of the wave function, the counter-clockwise/clockwise spatial evolution in the ring of the input state $\ket{\psi_{in}}$ at $\theta=0$ is given by $\Gamma_\pm (\theta) \ket{\psi_{in}}$ with
\begin{equation}
\Gamma_\pm(\theta) = U_z(\theta) e^{-i \Pi_\pm (l_z-\frac{\sigma_z}{2})\Pi_\pm \theta + i K_\pm R\theta }\Pi_\pm
\end{equation}
\noindent where $\Pi_\pm$ is the projector onto the subspace spanned by the polarization vectors $\{\ket{\chi(\pm k_\sigma)}\}$ with $k_\sigma$ positive, and $K_\pm$ is the generator of the space evolution in a straight segment at $\theta=0$, and it can be expressed as $K_\pm = \pm \sum_{\sigma \sigma'} k_\sigma \left(X^{-1}_0\right)_{\sigma\sigma'}\ket{\chi(\pm k_\sigma)}\bra{\chi(\pm k_{\sigma'})}$ with $\left(X_0\right)_{\sigma\sigma'}=\braket{\chi(\pm k_\sigma)}{\chi(\pm k_{\sigma'})}$.

\begin{figure}
[ht!]
\includegraphics[width=0.99\columnwidth]{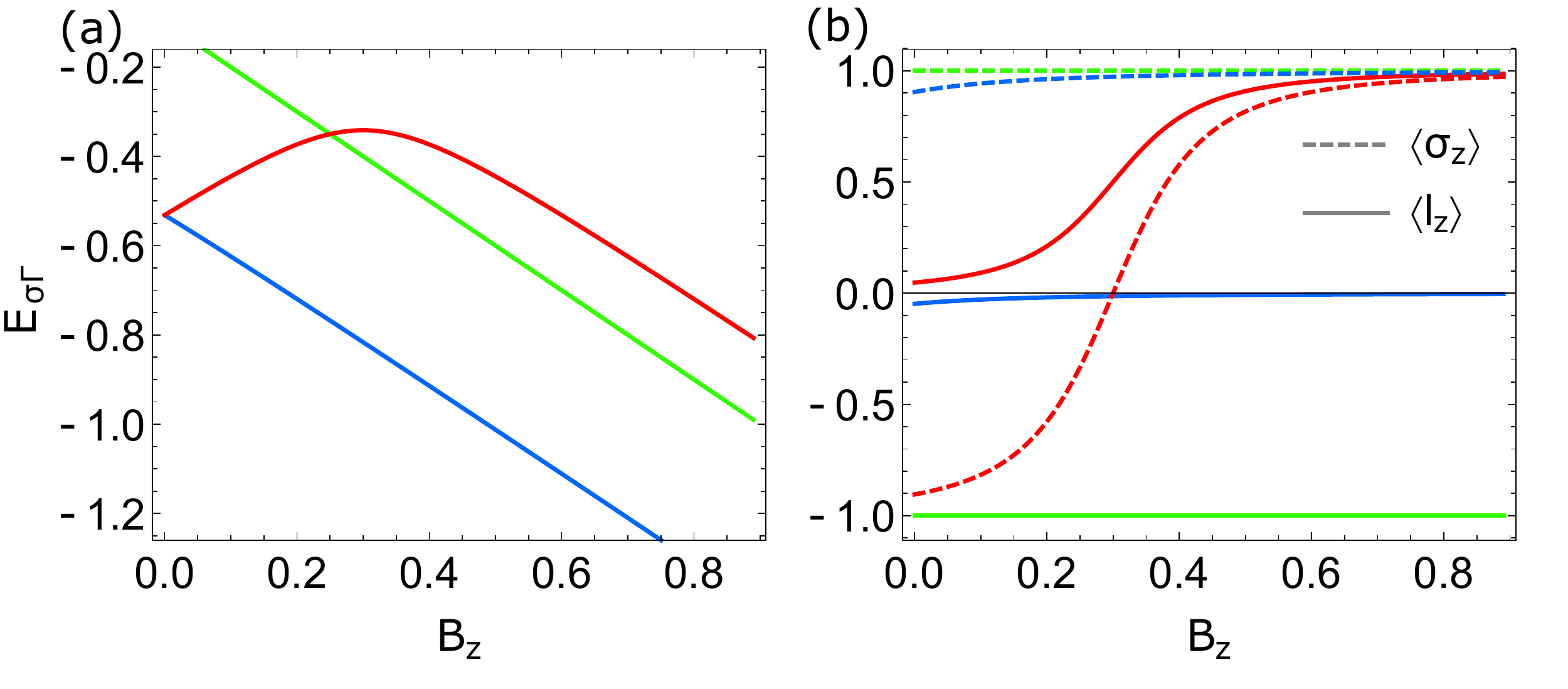}
\caption{(a) The energies $E_{\sigma,\Gamma}$ (a) (with $\sigma=1,2,3$ for the blue, red and green band, respectively), (b) spin and orbital orientations of the corresponding bands calculated at the $\Gamma$ point as a function of the magnetic field $B_z$. We observe that the orientations change keeping $\cos 2\pi (\langle l_z\rangle - \langle \frac{\sigma_z}{2}\rangle) =-1$. The $E_2$ band (red line) reaches its maximum at $B_z=-\Delta_t/2+\lambda_{SO}/2$ close to the position of the avoiding crossing with the band  $E_5$ thus influencing its spin-orbital features.
We use a representative set of parameters $\Delta_{t}=-0.5$, $\lambda=0.1$, $t_1=1.0$, and the energy scales are in unit of $t_2$.
}\label{fig:Energy_Gamma}
\end{figure}

\begin{figure}[ht!]
\includegraphics[width=0.99\columnwidth]{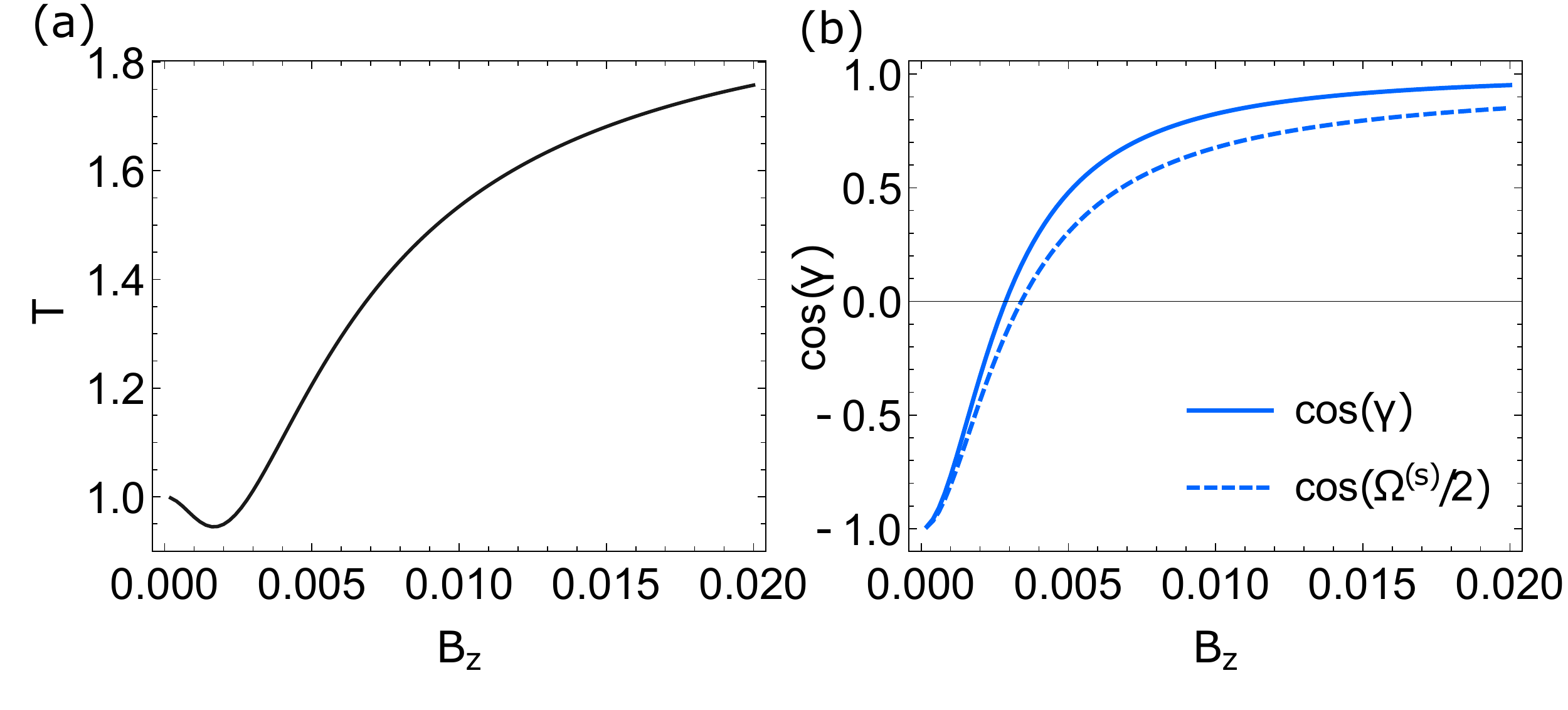}
\caption{(a) transmission coefficient and (b) geometric phase as a function of the transverse magnetic field $B_z$. The dashed line indicates the spin contribution to the geometric phase (b). We set $E_{in}=E_{1,\Gamma}|_{B=0}$, the ring radius $R=200$ (in unit of the atomic distances), $t_1=t_2=1$, $\Delta_{is}=0.2$, $\Delta_{t}=-0.5$,$\lambda=0.1$. We notice that in (b) the difference between the solid (total quantum phase) and the dashed (spin phase) curves is due to the orbital polarization.
}\label{fig:Gamma_point_BZ}
\end{figure}

Then, an input state $\ket{\phi^\pm_0}$  which is an eigenstate of $K_\pm R -\Pi_\pm\left(l_z-\frac{\sigma_z}{2}\right)\Pi_\pm$ with eigenvalue $\xi^\pm$, evolves in space as $ e^{i \xi^\pm \theta} U_z(\theta)\ket{\phi^\pm_0} $ in agreement with Eq.~\eqref{eq.sol circle}.

For an electron propagating in one of these channels within the ring, as it is commonly done for periodic systems, we get a quantum phase which can be separated as a sum of dynamical and geometric phases.
Since the spatial evolution is generated only by dipole terms, the AA  phase can be expressed as the sum $\gamma = \Omega^{(s)}/2 - \Omega^{(l)}$ where  $\Omega^{(s)}$ and $\Omega^{(l)}$ are the solid angles swept on the Bloch spheres by the spin $\langle \boldsymbol \sigma \rangle $ and orbital $\langle \mathbf l \rangle $ orientations with $\langle ... \rangle$ the expectation value for a given eigenstate.

The AA phase is associated with interference phenomena in the transport (here we assume a mirror symmetric configuration of the leads coupled to the ring).
In the limit of low bias applied voltage, the differential conductance at the energy $E_{in}$  can be obtained by means of the Landauer approach, and reads $g = \frac{e^2}{h} T$, where the transmission coefficient $T$ for unpolarized leads 
can be written as
\begin{equation}
T= 1 + \frac{1}{M}\textrm{Re}\Tr{\Gamma^\dagger_-(-\pi)\Gamma_+(\pi)}
\end{equation}
\noindent with $M$ being the number of the counter-clockwise (clockwise) channels, respectively.

Hence, the transmission amplitude can be expressed in terms of the AA phases $\gamma_\sigma$ as
\begin{equation}
T=1+ \frac{1}{M}\sum_{\sigma \sigma'} \abs{\braket{\phi^+_{0,\sigma}}{\phi^-_{0,\sigma'}}}^2 \cos\pi\left( \xi^{+}_{\sigma}+\xi^{-}_{\sigma'}\right)
\end{equation}
\noindent where $ 2\pi\xi^{\pm}_{\sigma} =d^\pm_\sigma + \gamma_\sigma -\pi$, $d^\pm_\sigma$ being the dynamical phase.

{\it Transverse magnetic field. -- } Let us then consider the effects of the transverse magnetic field on the transport. $B_z$ splits the energy degenerate states at the $\Gamma$ point and tends to pin the spin along the $z$ direction. At $\Gamma$ the $xy$ state is the lowest occupied and is separated by the $(zx,yz)$ configurations due to the crystal field potential $\Delta_t$ and the SO coupling. Our analysis is concentrated on the regime of low electron density where only the lowest energy states are occupied nearby the $\Gamma$ point. A representative snapshot of the energy window for a given set of electronic parameters is shown in Fig. \ref{fig:Energy_Gamma}(a). The spin-orbital features close the $\Gamma$ point are examined in terms of the expectation values $\langle l_z \rangle$ and $\langle \sigma_z \rangle$ in Fig. \ref{fig:Energy_Gamma}(b). 
The $xy$ lowest-energy state with opposite spin polarized configurations are Zeeman split and the separation increases until an avoiding level crossing at large magnetic field occurs (see Fig. 1 (a)).
The energy profile at the $\Gamma$ point allows to evaluate the electron transport for different spin and orbital channels.

When $E_{in}$ crosses only the band $E_{1}(k)$ at two points with moments $\pm k_1$, the AA phase uniquely contributes to the interference through the relation
\begin{equation}
T = 1 + \abs{\braket{\chi(k_1)}{\chi(-k_1)}}^2 \cos(\gamma)
\end{equation}
\noindent and the geometric phase and the transmission coefficient smoothly change with $B_z$ (see Fig.~\ref{fig:Gamma_point_BZ}).
We note that the interference is damped at small field by the almost vanishing overlap $\braket{\chi(k_1)}{\chi(-k_1)}\approx 0$.

\begin{figure}
[ht!]
\includegraphics[width=0.99\columnwidth]{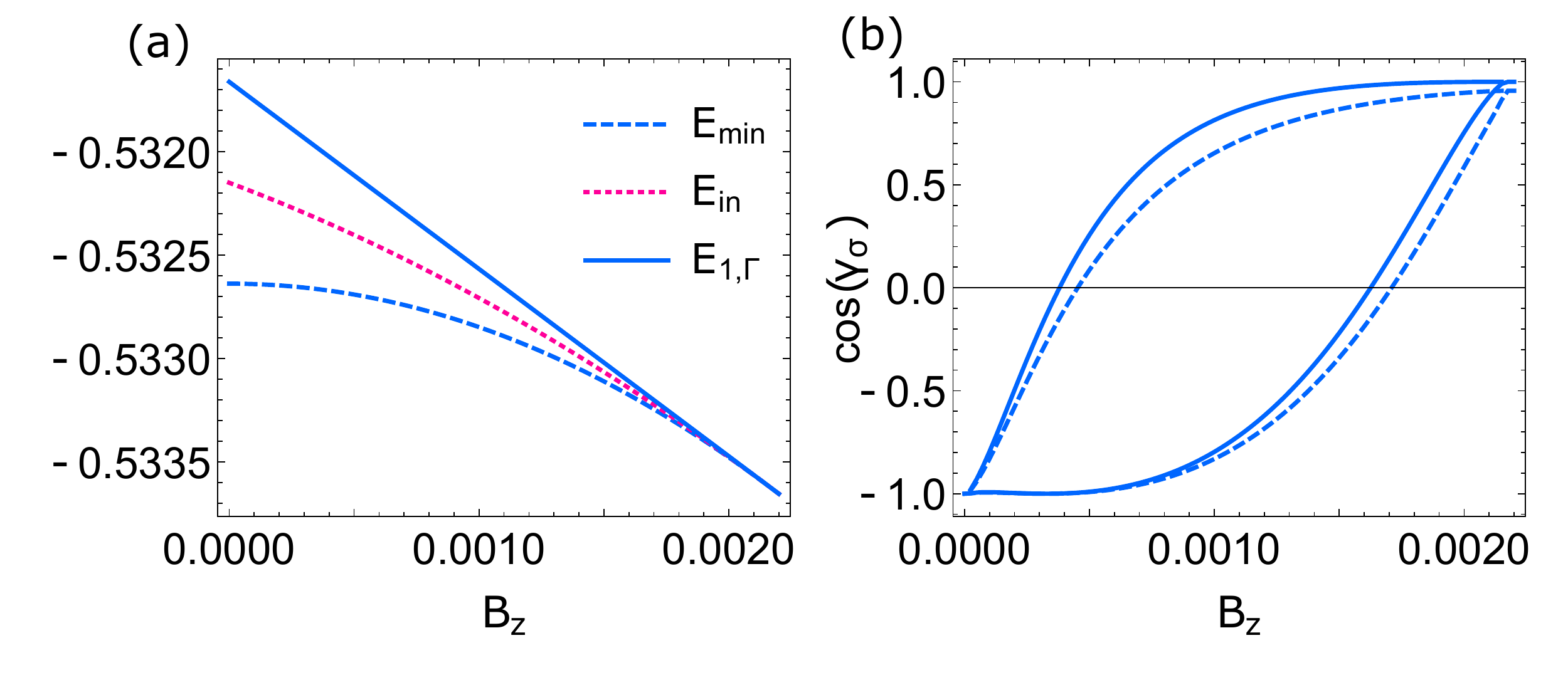}
\caption{We display (a) minimum $E_{min}$ of the band energy and the energies $E_{1,\Gamma}$ of the lowest band $E_1(k)$ at the $\Gamma$ point as a function of the transverse magnetic field $B_z$. There are two minima for small enough $B_z$ such that it is possible to obtain four crossings. In (b) we report the cosine of the AA phase for the total (solid) and spin (dotted) contributions. We consider $E_{in}$ (dotted line) equal to the average value of $E_{min}$ and $E_{1,\Gamma}$ and we set $R=200$, $t_1=t_2=1$, $\Delta_{is}=0.2$, $\Delta_{t}=-0.5$ and $\lambda=0.1$.
}\label{fig:low_BZ}
\end{figure}

Furthermore, the lowest band $E_{1}(k)$ displays two minima for small enough  $B_z$, which will merge by increasing $B_Z$ (see Fig.~\ref{fig:low_BZ}(a)).  By considering $E_{in}$ which crosses the band $E_{1}(k)$  at four points with momenta $\pm k_\alpha$ with $\alpha=1,2$, we find that the propagation occurs only in the clockwise sense (for the given choice of the field orientation).
This case corresponds to a chiral transmission which is related to the presence of skin modes having a topological origin~\cite{gong18}. More specifically, the transmission can be also described in terms of a non-hermitian tight-binding model and in this case for periodic boundary conditions the complex energies form two symmetrically related loops with winding number equal to $\pm 1$ (see Appendix).

\begin{figure}
[ht!]
\includegraphics[width=.99\columnwidth]{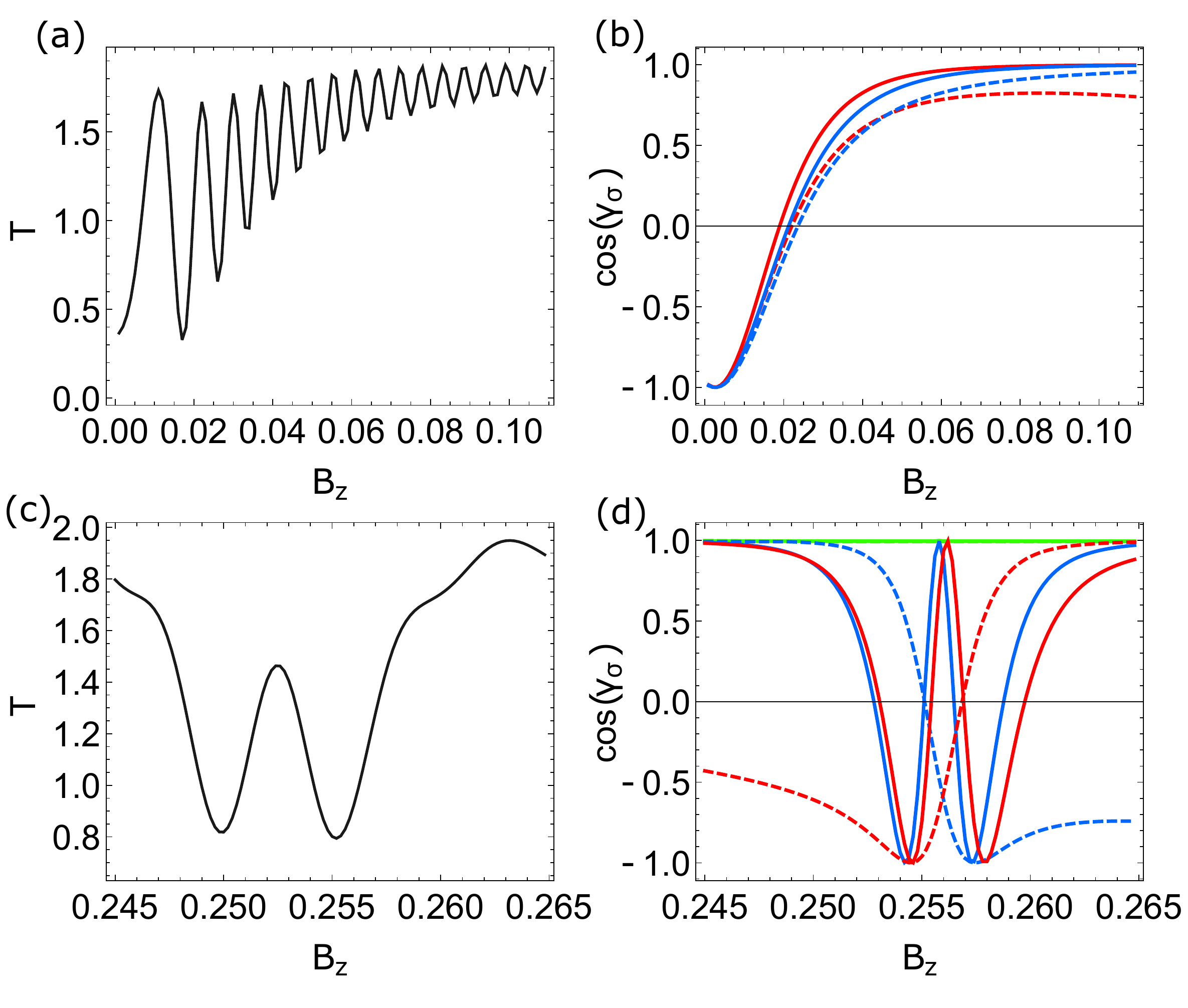}
\caption{(a) transmission coefficient and (b) geometric phase as a function of the magnetic field $B_z$. We consider $E_{in}=-0.3$, $R=200$, $t_1=t_2=1$, $\Delta_{is}=0.2$, $\Delta_{t}=-0.5$,$\lambda=0.1$. In (c) and (d) we report the behaviour in the high field regime for the transmission and the cosine of the AA phase, respectively. The colors refers to the electronic channels corresponding to the bands in Fig. 1. The dashed (solid) lines in (b) and (d) refer to the total (spin) phase while the difference between them highlights the role of the orbital contribution to the phase. We notice that approaching the regime of high field, $B_z \sim 0.250$, where the $xy$ configuration gets close to spin-orbital polarized state due to the superposition of $(xz,yz)$, the orbital contribution becomes significantly relevant and the transmission can have large amplitude modulations.
}\label{fig:large}
\end{figure}

A different behavior is displayed when the energy $E_{in}$ crosses more bands. When the two bands $E_{1,2}(k)$ are crossed, the presence of the magnetic field gives a separation of the two AA phases i.e. $\gamma_1\neq - \gamma_2$, which will contribute to the non-periodic oscillations in the transmission coefficient $T$ (see Fig.~\ref{fig:large}).

By increasing $B_z$ the crossings with the $E_2$ band get closer to the $\Gamma$ point, and the spin and orbital solid angles tend to balance each other giving $\cos \gamma_\sigma \approx 1$ with a damping of the oscillations in the transmission coefficient $T$.

When $B_z$ is large enough an avoided level crossing occurs at the $\Gamma$ point and the saturation of the phase $\gamma_\sigma$ is suddenly broken, with a sharp change in the transmission.

We observe that for very large $B_z$ the spin are pinned to the field direction, from which $\Omega^{(s)}_\sigma \to 0$ and a non zero phase $\gamma_\sigma$ come only from the orbital contribution $\Omega^{(l)}_\sigma$.

Concerning the observability of the outcomes of the analysis, we point out that for LaAlO$_3$-SrTiO$_3$ 2DEGs or at the surface of SrTiO$_3$, for the typical energy scales associated to the model Hamiltonian, the microscopic parameters of our interest assumes the values $\Delta_t \sim 50-100$ meV, $\Delta_{is}\sim 20$ meV, $\lambda_{SO}\sim 10$ meV, $t\sim 200-300$ meV,\cite{khalsa13,zhong,zabaleta,salluzzo} indicating that the regime at very large magnetic fields is inaccessible in laboratory. An alternative way to achieve such magnetic field amplitudes is to exploit the possibility of engineering interfaces with magnetic layers in proximity of the 2DEG \cite{salluzzoETO}: through the magnetic exchange, it will be thus possible to bring the 2DEG in a region of large spin and orbital polarizations. On the other hand, we find that an arbitrarily small field at which the saturation of the phase $\gamma_\sigma$ is broken can be achieved by lowering $\Delta_t$ or $\lambda_{SO}$ and tuning the parameters, thus indicating that the manipulation of the crystalline distortions, e.g. via strain, can be a way to observe the high field behavior by means of external magnetic fields.

{\it In-plane magnetic field. --} We proceed by considering an in-plane field $\mathbf B = B_x \hat x$ which preserves the symmetry under a wire-plane reflection $R_{xy}$, with a polygonal wire circumscribed in the circle. Differently to the transverse magnetic field case, if there is one counter-clockwise propagation channel in each side, the geometric phase is quantized i.e. $\cos\gamma^{(N)} = \pm 1 $.

\begin{figure}
[th!]
\includegraphics[width=0.99\columnwidth]{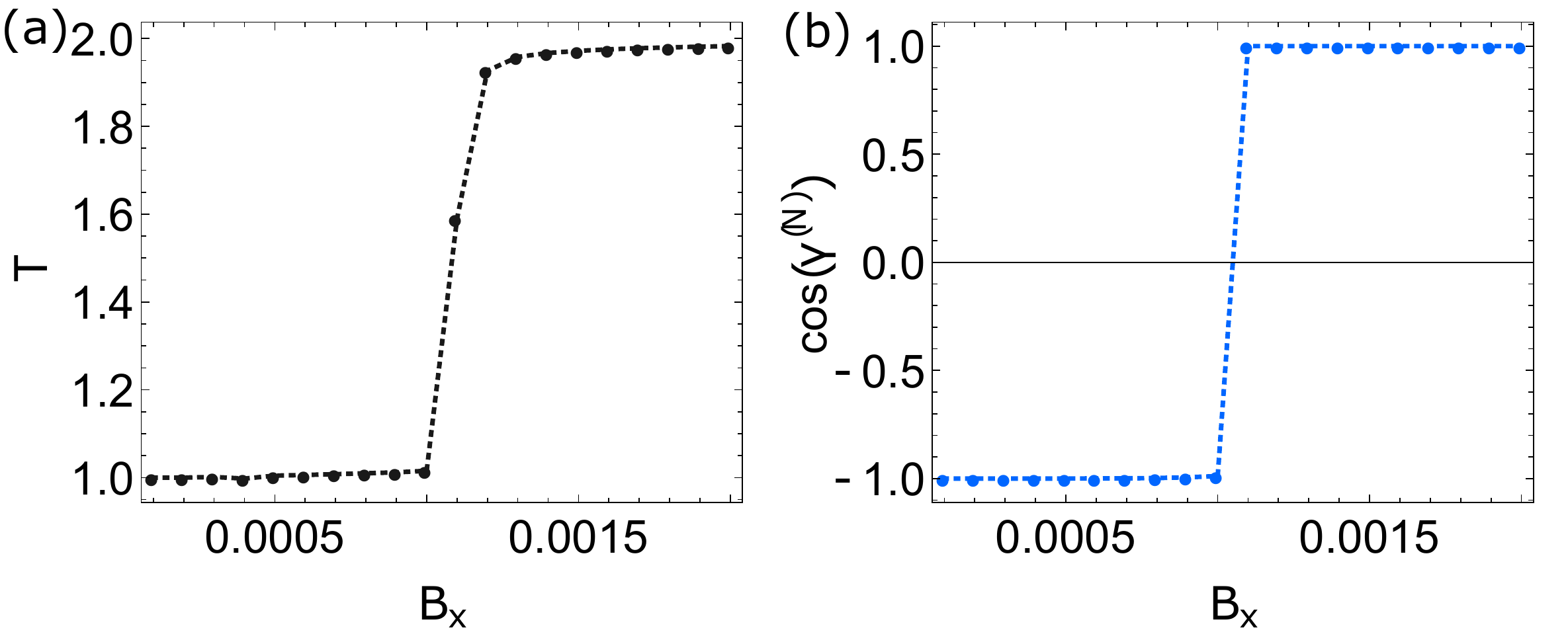}
\caption{ (a) transmission coefficient $T$ and (b) cosine of geometric phase as a function of the in-plane magnetic field $B_x$. For the computation we consider a polygonal wire of $N=500$ sides circumscribing the ring, and the microscopic parameters are $E_{in}=E_{1,\Gamma|_{B=0}}$, $R=200$,$t_1=t_2=1$, $\Delta_{is}=0.2$, $\Delta_{t}=-0.5$,$\lambda=0.1$.
}\label{fig:trans_BX}
\end{figure}

This can be understood by considering the orientations $\langle\boldsymbol \sigma\rangle$ and $\langle \mathbf l\rangle$ which point to the equators of the respective Bloch spheres, due to the planar reflection symmetry. The jump occurs when the spin orientation becomes parallel with respect to its initial orientation at $\theta=0$ in the space evolution so that it does not acquire a phase along the propagation.
We note that the interference effect even when $\cos\gamma^{(N)}=-1$ are strongly damped from which it results that $T\approx1$.

{\it Conclusions--} To wrap up, we have analysed the consequences of Zeeman magnetic field in the electron transmission through a quantum ring described by a model that can be relevant for nanochannels at the 2DEGs oxide interface, thus charcterized by local spin-orbital couplings arising from $d$-orbitals and orbital Rashba interactions.
The application of the field can lead to different transport regimes. For the case of only one counter-clockwise channel the AA phase is the unique source of interference in the transmission and the orbital phase is weakly contributing. Conversely, when more channels are considered there is also a dynamical part to include, and the coefficient of transmission shows non-periodic oscillations as a function of the magnetic field amplitude.
Furthermore, we find a chiral transmission at low energies, close to the bottom of the band dispersion, for small magnetic field strengths. We show that the contributions coming from the orbital polarizations remains small with respect to those arising from the spin precession at low magnetic field. By increasing the field amplitude a saturation of the AA phase towards zero value can occur, with sharp transitions and rapid variations when the field is strong enough and allows for orbital mixing.
The behavior drastically changes for an in-plane magnetic field, allowing a quantization of the AA phase due to the planar symmetry, with a transmission which sharply depends on the field. We observe that such quantization is only related to the spin state and thus it can be also achieved for the case of a single-mode semiconducting ring with Rashba interaction \cite{saarikoski}.

\section{Appendix}

The transmission in the circle can be also described with the help of a tight-binding model

\begin{equation*}
H=\sum_{\alpha \beta n} \left( T_{+, n+1,n}^{\alpha \beta} c^{\dagger}_{\alpha, n+1}  c_{\beta, n} + T_{-, n-1,n}^{\alpha \beta} c^{\dagger}_{\alpha, n-1}  c_{\beta, n} \right)
\end{equation*}

\noindent where we consider $N$ sites in the circle which are equidistant, i.e. at the vertexes of a polygon with $N$ sides circumscribed in the circle, $c^\dagger_{\alpha n}$ creates an electron in the spin-orbital state labeled with $\alpha$ at the vertex $n$, and $T_{\pm, n\pm1,n}^{\alpha,\beta}$ are the amplitudes of transmission in the $n\pm1$-th arc for an electron in the state $\ket{\beta}$ from the vertex $n$ to the state $\ket{\alpha}$ at the vertex $n\pm1$.
We note that the amplitudes for the selected states are non-homogenous so that we take in account the polarization states $\ket{\alpha,n}= U_z(n 2\pi /N) \ket{\alpha} $, thus obtaining a homogenous amplitude for transmission $\tilde T_{\pm}^{\alpha \beta}$. The effective Hamiltonian reads

\begin{equation*}
H=\sum_{\alpha \beta n} \left( \tilde T_{+}^{\alpha \beta} \tilde c^{\dagger}_{\alpha, n+1}  \tilde c_{\beta, n} + \tilde T_{-}^{\alpha \beta} \tilde c^{\dagger}_{\alpha, n-1} \tilde c_{\beta, n} \right)
\end{equation*}

\noindent For periodic boundary conditions one can perform a Fourier transform $ \tilde c_{\alpha n} = \frac{1}{\sqrt{L}}\sum_k e^{-i k n} \gamma_{k,\alpha}$ obtaining the model $H = \sum_{\alpha \beta k} H_{\alpha \beta}(k) \gamma^{\dagger}_{\alpha, k} \gamma_{\beta, k}$ with $H(k)= \tilde T_+ e^{ik}+\tilde T_- e^{-ik}$.
We find that, in the regime of the chiral transmission, the two eigenvalues $E_\sigma (k)$ of $H(k)$ draw two loops with winding number $\pm 1$ in the complex plane by changing $k$, which are related by an inversion transformation, showing a bulk-edge correspondence. This argument indicates that the chiral transmission can have topological features related with its intrinsic non-hermitian character.

\end{document}